\RequirePackage{etoolbox}
\patchcmd{\bibliographystyle}{#1}{rsc}{}{} 
\documentclass[sn-nature]{sn-jnl}

\usepackage{graphicx}%
\usepackage{multirow}%
\usepackage{amsmath,amssymb,amsfonts}%
\usepackage{amsthm}%
\usepackage{mathrsfs}%
\usepackage[title]{appendix}%
\usepackage{xcolor}%
\usepackage{textcomp}%
\usepackage{manyfoot}%
\usepackage{booktabs}%
\usepackage{algorithm}%
\usepackage{algorithmicx}%
\usepackage{algpseudocode}%
\usepackage{listings}%
\usepackage{lineno}	
\usepackage{soul}

\usepackage{setspace}

\begin{document}

\title[Article Title]{\textcolor{black}{Higher-order Hall response arises from octupole order and scalar spin chirality in a noncollinear antiferromagnet}}


\author[1]{\fnm{Adithya} \sur{Rajan}}

\author[1,2]{\fnm{Tom G.} \sur{Saunderson}}

\author[1,2,3]{\fnm{Fabian R.} \sur{Lux}}

\author[4,5]{\fnm{Roc\'io Yanes} \sur{D\'iaz}}

\author[6]{\fnm{Hasan M.} \sur{Abdullah}}

\author[1]{\fnm{Arnab} \sur{Bose}}

\author[1]{\fnm{Beatrice} \sur{Bednarz}}

\author[1,7]{\fnm{Jun-Young} \sur{Kim}}

\author[1,2]{\fnm{Dongwook} \sur{Go}}

\author[8]{\fnm{Tetsuya} \sur{Hajiri}}

\author[6]{\fnm{Gokaran} \sur{Shukla}}

\author[1]{\fnm{Olena} \sur{Gomonay}}

\author[9]{\fnm{Yugui} \sur{Yao}}

\author[9]{\fnm{Wanxiang} \sur{Feng}}

\author[8]{\fnm{Hidefumi} \sur{Asano}}

\author[6]{\fnm{Udo} \sur{Schwingenschl\"ogl}}

\author[4,5]{\fnm{Luis} \sur{L\'opez-D\'iaz}}

\author[1]{\fnm{Jairo} \sur{Sinova}}

\author[1]{\fnm{Gerhard} \sur{Jakob}}

\author[1,2]{\fnm{Yuriy} \sur{Mokrousov}}

\author[10]{\fnm{Aur\'elien} \sur{Manchon}}

\author*[1,11]{\fnm{Mathias} \sur{Kl\"aui}}\email{klaeui@uni-mainz.de}

\affil[1]{\orgdiv{Institute of Physics}, \orgname{Johannes Gutenberg-University Mainz}, \orgaddress{\street{Staudingerweg 7}, \city{Mainz}, \postcode{55128}, \country{Germany}}}

\affil[2]{\orgdiv{Peter Gr\"unberg Institut and Institute for Advanced Simulation}, \orgname{Forschungszentrum J\"ulich and JARA}, \orgaddress{\city{J\"ulich}, \postcode{52425}, \country{Germany}}}

\affil[3]{\orgdiv{Department of Physics}, \orgname{Yeshiva University}, \orgaddress{, \city{New York}, \state{NY}, \country{USA}}}

\affil[4]{\orgdiv{Department of Applied Physics}, \orgname{Universidad de Salamanca}, \orgaddress{\street{Plaza de la Merced}, \city{Salamanca}, \postcode{37008}, \country{Spain}}}

\affil[5]{\orgdiv{Unidad de Excelencia en Luz y Materia Estructuradas (LUMES)}, \orgname{Universidad de Salamanca}, \city{Salamanca}, \postcode{37008}, \country{Spain}}

\affil[6]{\orgdiv{Physical Science and Engineering Division}, \orgname{King Abdullah University of Science and Technology}, \orgaddress{\city{Thuwal}, \postcode{23955-6900}, \country{Saudi Arabia}}}

\affil[7]{\orgdiv{Institute of Materials Research and Engineering (IMRE)}, \orgname{Agency for Science, Technology and Research (A*STAR)}, \orgaddress{\city{Singapore}, \postcode{138634}, \country{Singapore}}}

\affil[8]{\orgdiv{Department of Materials Physics}, \orgname{Nagoya University}, \orgaddress{\city{Nagoya}, \postcode{464-8603}, \country{Japan}}}

\affil[9]{\orgname{Centre for Quantum Physics, Key Laboratory of Advanced Optoelectronic Quantum Architecture and Measurement (MOE), School of Physics, Beijing Institute of Technology, Beijing 100081}, \orgaddress{\city{Beijing}, \postcode{100081}, \country{China}}}

\affil[10]{\orgname{Aix-Marseille Universit\'e, CNRS, CINaM}, \orgaddress{\city{Marseille}, \country{France}}}

\affil[11]{\orgdiv{Centre for Quantum Spintronics, Department of Physics}, \orgname{Norwegian University of Science and Technology}, \orgaddress{\city{Trondheim}, \postcode{7491}, \country{Norway}}}




\abstract{\textcolor{black}{Noncollinear antiferromagnets can generate a transverse electrical response known as the anomalous Hall effect, even though they possess almost no net magnetization. The microscopic origin of this behaviour, however, has remained unclear because conventional measurement geometries mix different contributions to the measured response. Here, we show that applying magnetic fields in selected in-plane directions allows us to disentangle the mechanisms underlying the Hall effect in a representative noncollinear antiferromagnet. By suppressing any dipole-related signal, we isolate a purely octupole-driven Hall response that exhibits a characteristic three-fold angular symmetry. At low magnetic fields, we further observe an additional Hall-like contribution that arises from the scalar spin chirality associated with noncoplanar spin textures. Combining symmetry analysis, first-principles calculations, and transport measurements, we reveal that octupole order, dipole moments, and chirality coexist and contribute in distinct field regimes. These findings establish a framework for identifying and controlling complex magnetic order parameters for spintronic applications.}}




\maketitle


\section{Introduction}\label{sec1}


Broken time reversal symmetry ($\mathcal{T}$) and its interplay with the spin-orbit coupling (SOC) result in the transverse flow of electrons generating an anomalous Hall voltage, a signature that distinguishes a ferromagnet (FM) from conventional antiferromagnets (AFM) \cite{Nagaosa2010,Smejkal2022b}. 
However, it has been predicted that a certain class of AFMs with noncollinear spin textures can lead to an unusual type of anomalous Hall effect (AHE) that can exist even in the absence of a net magnetization due to spin-lattice coupling \cite{Chen2014a,Kubler2014}. 
Whilst in noncollinear antiferromagnets (NC-AFM) \textcolor{black}{it has been} shown that their weak magnetic moment is insufficient to account for the large AHE \cite{You2019,Iwaki2020,Takeuchi2021,Kiyohara2016,Nakatsuji2015}, further work to experimentally substantiate the \textcolor{black}{fundamental mechanism} responsible for the \textcolor{black}{observed AHE signals} \textcolor{black}{is missing.} This is because \textcolor{black}{most studies of the AHE have} been performed with the electrical measurements in plane while driving a magnetic field out of the plane, entangling the magnetization \textcolor{black}{signal} to any novel contribution to the AHE.
\textcolor{black}{In order to understand the origin of the AHE signal, one must go beyond the perpendicular field geometry and apply the field in different directions in space to identify the symmetry, and from this the mechanism leading to the AHE signal.} 
Here we show the full dependence of the AHE in the NC-AFM Mn$_3$Ni$_{0.35}$Cu$_{0.65}$N ($\mathrm{Mn_3NiCuN}$) when the magnetic field is swept \textcolor{black}{not only out-of-plane but also} in the plane.
\textcolor{black}{The in-plane field} experimental configuration, by construction, does not allow for the conventional dipole (magnetization) component of the AHE signal to contribute. We can therefore show, using a variety of theoretical techniques, that at high fields the in-plane AHE comes purely from the octupole moment and, surprisingly, \textcolor{black}{an additional} topological Hall-like (THE) signal \textcolor{black}{occurs} at low fields, consistent with prior evidence of scalar chirality-driven transport in the same compound \cite{Bose2025SOT}.
Our results demonstrate that in NC-AFMs one must expand beyond the dipole contribution to the AHE to explain their rich, coexisting orders. 
Such coexisting orders go beyond the magnetization dynamics achievable in conventional FMs and AFMs. Harnessing these coexisting orders \textcolor{black}{may enable future exploration of antiferromagnetic transport phenomena}. Details of the spin–orbit torque behavior in Mn$_3$Ni$_{0.35}$Cu$_{0.65}$N are discussed in a recent complementary study~\cite{Bose2025SOT}.

\begin{figure}
\centering
\includegraphics*[width=0.75\linewidth,clip]{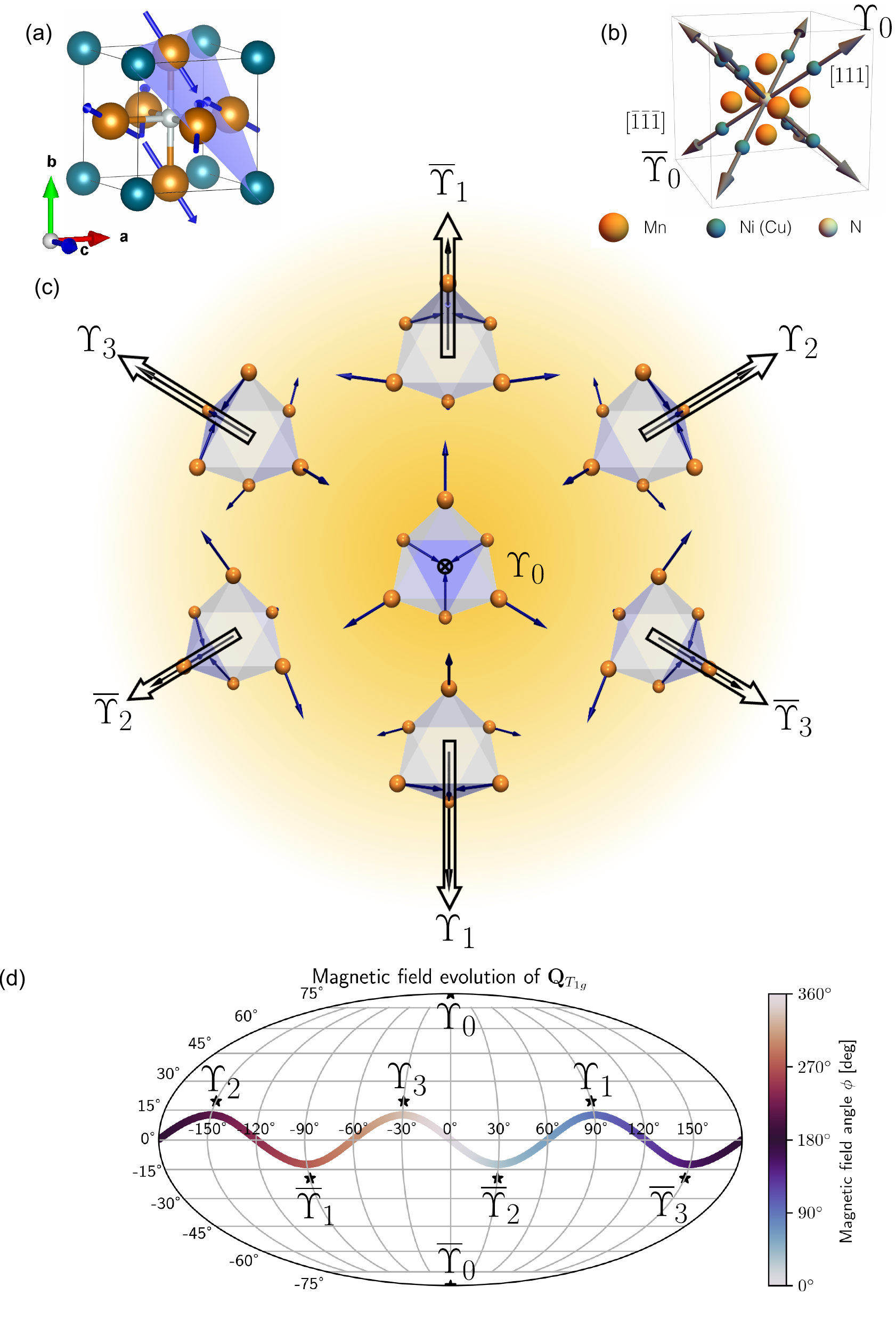} \\
\caption{\textbf{\textcolor{black}{Octupole structure and field-induced spin configurations in a noncollinear antiferromagnet.}}
(a) Crystal structure showing Mn atoms (orange), Ni/Cu atoms (blue), and N atoms (off-white). Blue arrows denote the coplanar 120° spin arrangement in the (111) kagome plane.
(b) Schematic representation of the eight symmetry-allowed octupole components, including the two out-of-plane projections labelled $\Upsilon_{0}$ and $\overline{\Upsilon}_{0}$, aligned with the [111] and $[\overline{1}\,\overline{1}\,\overline{1}]$ axes, respectively.
(c) Spin configurations and corresponding octupole vectors when magnetic fields are applied along selected crystallographic directions within the (111) plane. Each labelled direction matches an in-plane projection of an octupole component. The central configuration illustrates the case of an out-of-plane field probing the dominant octupole component.
(d) Calculated evolution of the octupole vector $\vec{Q}_{T_{1g}}$ in spherical coordinates as the in-plane magnetic field is rotated. The out-of-plane projection displays a 120$^\circ$ periodicity, which determines the expected angular dependence of the AHE.
}
\label{fig:1}
\end{figure}


The antisymmetric AHE in the NC-AFM arises from the octupole moment produced by the noncollinear frustrated spin structure in the kagome plane \cite{Chen2014a,Suzuki2017} which is in the (111) plane in the case of Mn$_3$NiCuN, the material we are investigating in this work (Fig.~\ref{fig:1}(a)). This can be visualized as an emergent octupole moment \cite{Kimata2021,Johnson2022,Suzuki2017,Lux2020,Gomonaj1989,Gomonaj1990a,Gurung2019a}, $\vec{Q}_{T_{1g}}$ (the first term in equation (\ref{Eq:AHE})) that purely originates from the coplanar orientation of the spin texture lying in the kagome plane (Fig.~\ref{fig:1}(a)). In $\mathrm{Mn_3NiCuN}$, the direction of $\vec{Q}_{T_{1g}}$ points out of the kagome plane, along [111] direction. \textcolor{black}{Upon expanding the AHE signal in the local spins, beyond the conventional ferromagnetic case, the octupole and topological Hall effects \cite{Martin2008,Bruno2004,MacHida2007,Taguchi2001} emerge as higher order contributions to the Hall effect in noncollinear compounds.} Hence, the total contributions to the AHE are,
\begin{linenomath*} 
\begin{equation}\label{Eq:AHE}
  \sigma_{xy}  = \gamma_{\mathrm{Oct}} ~ \left[\vec{Q}_{T_{1g}}\right]_{z} + \gamma_{\mathrm{dip}} ~ \left[\vec{M}_{T_{1g}}\right]_z + \gamma_\mathrm{SSC} ~ \left[\sum_{ijk}\vec{S}_i\cdot(\vec{S}_j \times \vec{S}_k)\right],
\end{equation}
\end{linenomath*} 
where equation (\ref{Eq:AHE}) is obtained by using representation theory to derive the irreducible representations of $\mathrm{Mn_3NiCuN}$ as described in section `2 Representation theory' of the supplementary material \cite{Supplementary}. The second term of equation (\ref{Eq:AHE}) represents the conventional AHE, proportional to the net magnetic dipole moment, $\vec{M}_{T_{1g}}$, as observed in a regular ferromagnet such as Fe \cite{Nagaosa2010}, and that can arise in our system due to the canting of the spins upon the application of an external magnetic field. Generally, in this type of NC-AFM \cite{Zhao2019}, including our system (see Supplementary Fig.~4(a) in `1.1 Sample characterization' \cite{Supplementary}), the induced moment is very small \cite{Zhao2019} and therefore has a very small effect on the AHE. 

\textcolor{black}{Equation (\ref{Eq:AHE}) predicts that $\vec{Q}_{T_{1g}}$ produces the maximum AHE when the magnetic field is swept out of the plane while the electrical measurements are performed in the kagome plane. This arrangement is similar to the measurement of AHE in a regular ferromagnet (FM)} where the direction of the applied electric current ($J_{xx}$), voltage measurement ($V_{xy}$) and magnetic moment ($M_z$) are orthogonal to each other ($V_{xy} \propto (J_{xx}M_{z})$). \textcolor{black}{Alternatively, in NC-AFMs with a vanishing magnetic moment, $\vec{Q}_{T_{1g}}$ plays the role of a fictitious magnetization, akin to $M_z$ in ferromagnets. Nonetheless our symmetry analysis reveals a fundamental difference between the octupole driven AHE and the conventional dipole driven AHE: the magnetic octupole supports a specific finite AHE response when the external field is rotated \textit{in the plane} of the electrical measurements, whereas the dipole driven AHE remains unaffected.} Thus, in plane fields have to be used to reveal this contribution.

\textcolor{black}{The octupole vector} $\vec{Q}_{T_{1g}}$ hosts a total of eight poles in $\mathrm{Mn_3NiCuN}$ as shown with the grey vectors in Fig.~\ref{fig:1}(b), two of which point out of the plane, along the [111] direction and six others have projections both in the plane and out of the plane with 120$^{\circ}$ in-plane rotational symmetry enforced by the crystal structure (space group $Pm \overline{3} m$ (No. 221), point group $m \overline{3} m$ ($O_{h}$)). It is evident that when an external field is applied out-of-plane, along the [111] axis, one expects to obtain an AHE from the combination of octupole ($\vec{Q}_{T_{1g}}$) and dipole ($\vec{M}_{T_{1g}}$) contributions. However, our symmetry analysis additionally shows (see `2.6 Approximate ground state' \cite{Supplementary}) that when the in-plane applied magnetic field is collinear to the in-plane projections of any one of the six components of these $\vec{Q}_{T_{1g}}$ vectors, it immediately couples to the octupoles by \textcolor{black}{reorienting} the spin-configuration into one of the equivalent (111) planes as shown in Fig.~\ref{fig:1}(c). \textcolor{black}{This coupling provides direct access} to the out-of-plane projection of $\vec{Q}_{T_{1g}}$ which can lead to an AHE as per equation (\ref{Eq:AHE}) when electrical measurements are \textcolor{black}{performed} in the plane. Due to this, one expects a 120$^{\circ}$ angular dependence of the measured AHE as theoretically calculated in Fig.~\ref{fig:1}(d). We point out here that we plot all components of the $\vec{Q}_{T_{1g}}$ vector in Fig.~\ref{fig:1}(d) and find that \textcolor{black}{its dependence on the magnetic field resembles that of the magnetization, $\vec{M}_{T_{1g}}$, which illustrates its role as a fictitious magnetization; however, the key difference is that the octupole produces finite contributions to the in-plane AHE, unlike $\vec{M}_{T_{1g}}$.} This defines a distinct difference from the \textcolor{black}{conventional (dipole-driven)} AHE in a FM as the AHE is not allowed to occur when electrical measurements and applied magnetic field are coplanar.

In addition, our theoretical calculations suggest a third term in equation (\ref{Eq:AHE}) that predicts a nontrivial Hall effect originating from the scalar chirality of the spin textures. \textcolor{black}{Such a quantity was first discussed in high temperature superconductors \cite{Wen1989,Lee1992} when considering contributions to the Hall conductivity, whose underlying physics can similarly be attributed to the THE in a skyrmion \cite{Kimbell2022,Martin2008,Bruno2004,MacHida2007,Taguchi2001}}. This can be observed in the low field regime when the spins reorient, maximizing the scalar spin chirality.

\textcolor{black}{In this work, we reveal the microscopic origin of the AHE in a NC-AFM by systematically probing how the response evolves when magnetic fields are applied both out of the plane and along crystallographically defined in-plane directions. This approach allows us to separate any residual dipole contribution from the intrinsic octupole-driven signal and to identify the characteristic three-fold rotational symmetry associated with the octupole order. We further uncover an additional low-field Hall-like contribution that emerges when the spins form noncoplanar textures with finite scalar chirality. By combining symmetry analysis, first-principles calculations and magnetotransport measurements, we map how these different orders coexist and dominate in distinct field regimes. Our results establish a clear experimental and theoretical framework for interpreting Hall transport in noncollinear antiferromagnets and highlight the multi-component magnetic orders that can be harnessed for future spintronic functionality.}

\section{ Results}\label{sec2}

\begin{figure}
\centering
\includegraphics*[width=1.0\linewidth,clip]{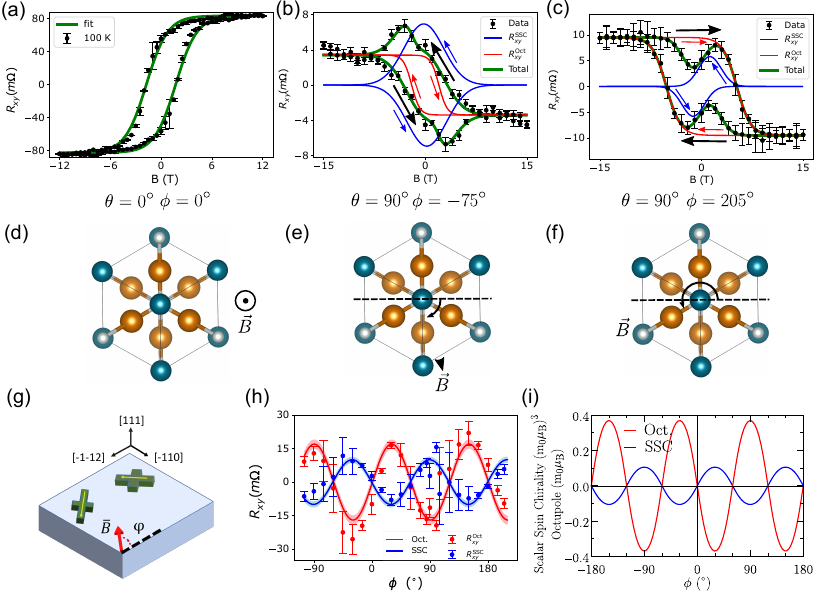} \\
\caption{ 
\textbf{\textcolor{black}{Angular evolution of octupole-driven and chirality-driven Hall responses.}}
(a–c) Transverse Hall resistance measured as a function of magnetic field applied (a) out of the kagome plane and (b–c) within the plane at two representative in-plane field angles. Black symbols show experimental data; green curves are fits to the combined model in Equation (2), consisting of an octupole-driven term (red curves) and a scalar-chirality-driven term (blue curves). Small arrows indicate the direction of hysteresis loop shift.
(d–f) Schematics of the magnetic-field directions corresponding to the measurements in (a–c).
(g) Device geometry used for magnetotransport measurements, indicating crystallographic axes and the orientation of current and voltage lines.
(h) Extracted amplitudes of the octupole-driven contribution $R^{\mathrm{Oct}}_{xy}$ (red) and the scalar-chirality contribution $R^{\mathrm{SSC}}_{xy}$ (black) as functions of in-plane angle. Both show an approximate 120° periodicity. Error bars represent the standard deviation of five repeated measurements.
(i) Symmetry-based theoretical calculation of scalar spin chirality, displaying the same 120° periodicity but with an opposite sign compared to the octupole contribution.
}

\label{fig:THE}
\end{figure}

To test these theoretical predictions, $\mathrm{Mn_3NiCuN}$ (111) thin film of 20 nm thickness are grown on an MgO (111) substrate, and then capped with a 3 nm thin Pt layer to prevent oxidation. Crystallographically aligned Mn$_3$NiCuN growth on MgO was confirmed by X-ray diffraction measurements with Cu K$_\alpha$ radiation (See Supplementary Figs.~s 1-3 in the Supplementary \cite{Supplementary}). After the thin film growth, Hall bar devices were patterned on the (111) kagome plane as shown in Fig.~\ref{fig:THE}(g). The details of the sample preparation can be found in the method section and in reference \cite{Zhao2019}. First, we verify the presence of the $\vec{Q}_{T_{1g}}$ component's contribution to the AHE signal by measuring the standard Hall effect in the kagome plane while sweeping the magnetic field out-of-plane field along the [111] direction (z-axis) at 100 K, well below the N\'eel temperature ($T_N \sim$ 200 K) \cite{Zhao2019}. The longitudinal resistivity of the film is $\sim150~\mu \Omega \mathrm{cm}^{-1}$ at a temperature 100 K. Fig.~\ref{fig:THE}(a) shows the measured AHE data for the z-field sweep that can be fitted by using \textcolor{black}{ $R_{xy} = R_{xy0} \tanh(a(B - B_c))$, where $R_{xy0}$ is the saturation amplitude, $a$ is a scaling factor and $B_c$ the coercive magnetic field strength}. We also verify that this AHE signal disappears above the N\'eel temperature (see Supplementary Fig.~4(b) in `1.1 Sample characterization' \cite{Supplementary}). This result is consistent within the framework of the octupole moment ($\vec{Q}_{T_{1g}}$) induced AHE (equation (\ref{Eq:AHE}), first term), but additionally a small magnetic moment is also present in our system (see Supplementary Fig.~4(a) in `1.1 Sample characterization' \cite{Supplementary}), meaning that the magnetic dipole component ($\vec{M}_{T_{1g}}$) also contributes to the AHE. The measured AHE is comparable to previous reports \textcolor{black}{in $\mathrm{Mn_3NiCuN}$ and $\mathrm{Mn_3NiN}$ but lower in magnitude than other NC-AFMs such as $\mathrm{Mn_3Ge}$ and $\mathrm{Mn_3Sn}$ \cite{Miki2020,Nakatsuji2015,Boldrin2019,Kiyohara2016}.}

Now we perform the measurements of transverse resistance on the (111) plane while sweeping the magnetic field in the plane along different in-plane crystallographic directions (Figs.~\ref{fig:THE}(b-c)). We observe two important features: (1) AHE-like signals, representing a step in the measured resistance, and (2) topological Hall-like (THE) signals, which are additional features only appearing in the low field regime which we will discuss later. Our data can be fitted well to equation (\ref{Eq:fit}),
\begin{linenomath*} 
\begin{equation}\label{Eq:fit}
  R_{xy}(B) = R^\mathrm{Oct}_{xy} \tanh(a(B - B_\mathrm{c})) + R^\mathrm{SSC}_{xy} \exp\left(-c(B-B_0)^2\right),
\end{equation}
\end{linenomath*} 
where $R^\mathrm{Oct}_{xy}$ represents the maximum contribution coming from the octupole moment, and $R^\mathrm{SSC}_{xy}$ the maximum contribution of the scalar spin chirality, $B$ is the magnitude of the magnetic field, $B_c$ is the coercivity, \textcolor{black}{$a$ is related to the slope of $R^\mathrm{Oct}_{xy}$ switching,} and c is related to the width of the scalar spin chirality signal. 

The observation of AHE-like signal in such coplanar geometry is a nontrivial finding as \textcolor{black}{all but the highest coplanar anisotropies will not manifest a dipole contribution of this form}. \textcolor{black}{As such coplanar anisotropies do not exist in this material,} the observed signal implies that applied magnetic fields can access one of the components of $\vec{Q}_{T_{1g}}$ depending on the direction of the magnetic field-sweep with respect to the crystal axis (Figs.~1(c-d)). \textcolor{black}{This is} further evident following the red curve in Fig.~\ref{fig:THE}(h) that exhibits a roughly 120$^{\circ}$ angular dependence, consistent with our theoretical predictions (Fig.~\ref{fig:1}). The error-bars in the data points represent the standard deviation for five different measurements.

In these experiments surprising additional features appear at low field values which can be fitted by a Gaussian function that changes sign for positive and negative applied field \textcolor{black}{cycling direction}. 
Features like these are often attributed to THE as they occur in the presence of skyrmions whose nontrivial topology can induce an emergent effective magnetic field \cite{Kimbell2022,Bruno2004}. The spin textures in our system exhibit no such topology \cite{Kimbell2022}, however it is possible that in the low field regime the spins rotate out of their coplanar orientation and maximize the scalar spin chirality ($\vec{S}_i\cdot(\vec{S}_j\times\vec{S}_k)$) \cite{Martin2008} thereby producing THE-like signals as predicted by the third term in equation (\ref{Eq:AHE}). \textcolor{black}{A consequence of this is that for certain $\phi$ where the THE-like component is larger than the $\vec{Q}_{T_{1g}}$ component, $R_{xy}$ switches before B changes polarity as shown in Fig.~\ref{fig:THE}(b) for $\phi = -75^{\circ}$. This would be thermodynamically prohibited if the AHE has contributions only from $\vec{Q}_{T_{1g}}$ or $\vec{M}_{T_{1g}}$.} 
To verify the source of the additional contribution, we compute the components of the scalar spin chirality by parameterizing a free energy expression of the irreducible representations with first principles calculations and solve a Landau-Lifshitz-Gilbert equation to provide the dependence of the scalar spin chirality ($\vec{S}_i\cdot(\vec{S}_j\times\vec{S}_k)$) and octupole ($\vec{Q}_{T_{1g}}$) moment as a function of in-plane magnetic field angle $\phi$. The simulation suggests that the noncoplanar spin-textures are possible and predicts that both the octupole signal (arising from $\vec{Q}_{T_{1g}}$) and the THE-like signal (arising from $\vec{S}_i\cdot(\vec{S}_j\times\vec{S}_k)$) possess an angular dependence with 120$^{\circ}$ rotational symmetry (Fig.~\ref{fig:THE}(i)). This result is qualitatively consistent with the  angular dependence of the both the signals in our experiment (Fig.~\ref{fig:THE}(h)) complete with their opposing signs.

\section{Discussion}\label{sec13}

Whilst attempts have been made to quantify the behavior of the complex spin structures present in NC-AFMs using simple pictures to explain \textcolor{black}{the sign change, for example, in the AHE} \cite{You2019,Iwaki2020,Kiyohara2016,Nakatsuji2015,Kimata2019b,Nan2020,Xie2022,Chen2023,Qin2023,Holanda2020,Takeuchi2021,Zhang2016,Reichlova2019,Rout2019}, our work provides a full understanding of the signal of the AHE as a function of the magnetic field. Notable exceptions to this rule include Song et. al. \cite{Song2022}, however they probed the symmetric \textit{planar} Hall effect and not the antisymmetric AHE. Additionally work by Ghosh et. al. \cite{Ghosh2023} provided \textcolor{black}{an argument} for an alternative origin for the THE-like signal in Nd$_2$Ir$_2$O$_7$, however this origin \textcolor{black}{suggests} the presence of Weyl nodes arising from broken inversion symmetry which is not present in our system. 

\textcolor{black}{To rule out other sources of the measured signals, one needs to confirm the crystalline structure and orientation of the thin film. In particular, the orientation of the film's crystal structure needs to be checked to rule out the effects of other phases or orientations. We find that our thin films exhibit alignment with both the OOP and in-plane directions of the substrate (See section `1.1 Sample Characterization' in the Supplementary \cite{Supplementary}). However, the non-negligible lattice mismatch between the substrate and $\mathrm{Mn_3NiCuN}$ presents a situation where one does not expect a perfect epitaxial thin film alignment. While it is thus clear that qualitatively, the observed symmetries of the transport signal mirror the symmetry of the crystal structure with no other phases and orientations present, the small local variations of the crystallinity can result in the attenuation of the $R_{xy}$ in comparison to a fully crystalline bulk sample, for instance.} This is compounded by the current shunting effect due to the Pt layer on top, due to which a 20{\%} underestimation of $R_{xy}$ can be expected. However, this is merely a suppression of the overall signal and does not lead to any modification in the interpretation of the underlying effect. 
The Pt layer could potentially induce magnetoresistance effects at the interface due to spin accumulation and adds a degree of uncertainty into the interpretation of the data. However, the consistent 120$^{\circ}$ symmetry of both the AHE and SSC signal at two different devices and across a range of temperatures (See Supplementary Fig.~7 `1.2 Data processing' in the Supplementary \cite{Supplementary}) provides compelling evidence in support of our interpretation.
\textcolor{black}{To check the influence of Pt, additional measurements on a $\mathrm{Mn_3NiCuN}$ thin film without a capping Pt layer were performed (See Supplementary Fig.~8 `1.2 Data processing' in the Supplementary \cite{Supplementary}). The $R^\mathrm{Oct}_{xy}$ and $R^\mathrm{SSC}_{xy}$ show analogous $\phi$-dependence as for the stack shown in Fig.~\ref{fig:THE}(h) showing that any potential effects of the Pt capping layer are negligible for the purpose of our work.}

Our analysis of the underlying spin dynamics elucidates a clear interplay of coexisting orders within a NC-AFM, which sets this class of magnetic order apart from conventional ferromagnets and antiferromagnets. These collinear spin structures host only one order parameter apiece, the magnetization and the N\'eel vector, whereas NC-AFM exhibit further complexities that have direct consequences on the AHE that can only be understood through expansion in local spins beyond the ferromagnetic case. For $\mathrm{Mn_3NiCuN}$, and subsequently the whole class of $\mathrm{Mn_3XY}$ compounds, one can uncover the coexistence of three distinct orders: the magnetization, a quantity in part responsible for the conventional AHE in this NC-AFM, the octupole contribution, that provides both conventional and also in-plane AHE contributions and finally the THE-like component generated when the spins are noncoplanar. Although this particular material is merely a prototypical example of the vast array of other NC-AFM orders, we present a very clear paradigm shift in the way these materials are to be viewed: that they have multiple unconventional order parameters which behave like a magnetization, yet they affect the Hall transport in unique ways. \textcolor{black}{This procedure can also be applied to other classes of materials, such as Mn$_3$X compounds, which have already shown similar in-plane AHE signatures \cite{You2019}.} 

In summary, we systematically study the symmetry of the transverse Hall resistance in $\mathrm{Mn_3NiCuN}$ (111) by applying magnetic fields in the kagome plane along selected crystallographic directions. We find that whenever the external applied magnetic field is collinear to the in-plane projections of any one of the octupoles, an AHE can be detected even in the limit of vanishing magnetization. The six octupoles that have in-plane components lead to a 120$^{\circ}$ symmetric AHE signal even when magnetic field and electrical measurements are coplanar, a remarkable \textcolor{black}{contrast} from the conventional AHE in FMs. We observe THE-like features \textcolor{black}{and decompose the signal into octupole and THE-like components, finding the THE-like components to behave} with a similar 120$^{\circ}$ angular dependence as the octupole component while sweeping the magnetic field in the kagome plane. We attribute this effect to a scalar-chirality emerging during the spin-reorientation at low magnetic fields. Our experimental results are well supported by theoretical predictions. \textcolor{black}{Our work represents a key step in uncovering the complex, nontrivial phenomena in NC-AFMs, greatly enhancing the understanding of the octupole moments, and distilling the complex spin textures into clear, coexisting orders. This understanding opens the possibility to explore multiple means to harness NC-AFMs for novel spintronic technological applications. }

\backmatter

\subsection{Acknowledgement}

We acknowledge funding from King Abdullah University of Science and
Technology (KAUST) under award 2020-CRG8-4048 and 2024-CRG12-
6480. A.M. was supported by the Excellence Initiative of Aix-Marseille University
-A*MIdex, a French “Investissements d’Avenir program”. In addition,
the Deutsche Forschungsgemeinschaft (DFG, German Research Foundation)
$-$ Grant No. TRR 173/2 $-$ 268565370 Spin+X (projects A01, B02 and
A11) and the European Union’s Framework Programme for Research and
Innovation Horizon Europe under grant number 101226840 (ORBIS) and
grant number 101129641 (OBELIX)) are acknowledged. A.R., B.B., L.L.D.,
G.J. and M.K. acknowledge funding from the European Union’s Framework
Programme for Research and Innovation Horizon 2020 (2014$-$2020) under
the Marie Skłodowska-Curie Grant Agreement No. 860060 (ITN MagnEFi).
Y.M., J.S., W.F., and Y.Y. acknowledge funding under the National Natural
Science Foundation of China (Grant No.W2511003), the Joint Sino-German
Research Projects (Chinese GrantNo. 12061131002 and German Grant No.
1731/10-1), and the Sino-German Mobility Programme (Grant No. M-0142).
A.B. thanks the Alexander Von Humboldt Foundation for the postdoctoral
fellowship. R.Y.D. and L.L.-D. acknowledge support form project PID 2023-
150853NB-C31 funded by MICIU/AEI /10.13039/501100011033 and by
FEDER, UE. T.G.S., F.R.L., D.G. and Y.M. acknowledge the Jülich Supercomputing
Centre for providing computational resources under project
jiff40. T.H. and H.A. acknowledge funding from the Japan Society for the
Promotion of Science (KAKENHI Grant Nos. 20H02602 and 19K15445).

\subsection{Author Contributions}
A. Rajan performed the experimental measurements, T. G. Saunderson performed the first principles calculations and wrote the paper with A. Rajan and A. Bose. F. R. Lux performed the symmetry analysis, R. Yanes D\'iaz and L. L\'opez-D\'iaz performed atomistic spin dynamics simulations. H. M. Abdullah performed tight binding calculations and provided insights on the scalar spin chirality with G. Shukla, U. Schwingenschl\"ogl and A. Manchon. J.-Y. Kim assisted in the experimental discussions, T. Hajiri and H. Asano performed thin film sample deposition, B. Bednarz assisted with device fabrication. G. Jakob assisted in structural characterization and provided insights on experimental discussion. Y. Yao, W. Feng, Y. Mokrousov and D. Go provided insights in the transport from the first principles perspective, O. Gomonay and J. Sinova provided insights on the symmetry analysis. T. G. Saunderson, F. R. Lux and Yuriy Mokrousov coordinated the theory effort. M. Kl\"aui provided experimental insight and was the principal investigator, supervising the whole project.

\subsection{Corresponding author}
Correspondence to Mathias Kläui (klaeui@uni-mainz.de).

\textcolor{black}{\section{Competing interests}
The authors declare no competing interests.}

\section{Data availability}
\textcolor{black}{The raw and processed data that support the findings of this study} are available from the corresponding author upon reasonable request.

\section{Code availability}
\textcolor{black}{The codes used for analysing the data in this study} are available from the corresponding author upon reasonable request.

\section{Methods}\label{sec11}

\subsection{Material and Device fabrication}

On MgO (111) substrate, 20 nm Mn$_3$NiCuN thin film was deposited using reactive magnetron sputtering  at 375$^\circ$C substrate temperature under 2.0 Pa with 4\% N$_2$ + 96\% Ar gas mixtures. After growth, the sample was annealed in-situ at 500$^\circ$C under the same atmosphere as film growth. Please refer to \cite{Zhao2019} for further details on thin film deposition. A 3 nm Pt layer was deposited in-situ to prevent oxidation. 
After the thin film growth, \textcolor{black}{Hall bar devices with current line of width 10 $\mu$m and voltage line of width 3 $\mu$m}  were patterned using electron beam lithography, and the surrounding area removed using  Ar$^+$ ion etching.  

\subsection{Transport measurements}
The sample with patterned Hall bars was mounted on a standard PCB with Au contacts contacted to the Hall bars by wire bonding using TPT Hybond 572-40. The PCB with contacted sample was then mounted on attocube ANRv51/RES piezo-controlled rotatable sample holder which was inserted in a variable temperature cryostat from Oxford Instruments to perform magneto-transport measurements. Keithley 2400 was used as a current source, and the Hall voltage was measured using Keithley 2182A nano-voltmeter. 

\subsection{Data processing}
The acquired Hall signal was anti-symmetrized ($R_{xy}(B)= (V_{xy}(B) - V_{xy}(-B)/2I_{xx})$ to extract the AHE, and to remove any contributions from the longitudinal signal. The anomalous Hall signal was centered and the contribution linear with B corresponding to the ordinary Hall effect subtracted. The planar hall effect was also analyzed and \textcolor{black}{removed} as in Supplementary Fig.~5 and is plotted in Supplementary Fig.~6. \cite{Supplementary}. 

\subsection{Computational details}
We performed Density Functional Theory calculations for bulk Mn$_3$NiCuN using the experimental lattice constant $3.9012 \AA$ from Zhao \textit{et al} \cite{Zhao2019}. We performed calculations using the FLEUR code (for the program description, see https://www.flapw.de) which implements the full potential linear augmented plane wave method (FP-LAPW) \cite{Wimmer1981}, employing the generalized gradient approximation (GGA) \cite{Perdew1996}. For the symmetry analysis, full self consistencies
were obtained at multiple noncoplanar canting angles from the $\Upsilon_0$ spin configuration in Fig.~\ref{fig:1}(a). In order to quantify the AHE from first principles, we employed Wannier interpolation \cite{Freimuth2008, Pizzi2020} to efficiently compute the Berry curvature using the Kubo formalism \cite{Nagaosa2010}. Full details are provided in section `3 Numerical details' of the Supplemental material.

\subsection{LLM Disclosure}
Large language model assistance (ChatGPT) was used for language editing and formatting suggestions only. All scientific content, data interpretation, and conclusions are solely the authors’ work.

\providecommand*{\mcitethebibliography}{\thebibliography}
\csname @ifundefined\endcsname{endmcitethebibliography}
{\let\endmcitethebibliography\endthebibliography}{}

\newpage

\section*{\textcolor{black}{Figure Captions}}

\noindent \textbf{\textcolor{black}{Figure 1 | Octupole structure and field-induced spin configurations in a noncollinear antiferromagnet.}}
\textcolor{black}{(a) Crystal structure showing Mn atoms (orange), Ni/Cu atoms (blue), and N atoms (off-white). Blue arrows denote the coplanar 120° spin arrangement in the (111) kagome plane.
(b) Schematic representation of the eight symmetry-allowed octupole components, including the two out-of-plane projections labelled $\Upsilon_{0}$ and $\overline{\Upsilon}_{0}$, aligned with the [111] and $[\overline{1}\,\overline{1}\,\overline{1}]$ axes, respectively.
(c) Spin configurations and corresponding octupole vectors when magnetic fields are applied along selected crystallographic directions within the (111) plane. Each labelled direction matches an in-plane projection of an octupole component. The central configuration illustrates the case of an out-of-plane field probing the dominant octupole component.
(d) Calculated evolution of the octupole vector $\vec{Q}_{T_{1g}}$ in spherical coordinates as the in-plane magnetic field is rotated. The out-of-plane projection displays a 120$^\circ$ periodicity, which determines the expected angular dependence of the AHE.}

\bigskip

\noindent \textbf{\textcolor{black}{Figure 2 | Angular evolution of octupole-driven and chirality-driven Hall responses.}}
\textcolor{black}{(a–c) Transverse Hall resistance measured as a function of magnetic field applied (a) out of the kagome plane and (b–c) within the plane at two representative in-plane field angles. Black symbols show experimental data; green curves are fits to the combined model in Equation (2), consisting of an octupole-driven term (red curves) and a scalar-chirality-driven term (blue curves). Small arrows indicate the direction of hysteresis loop shift.
(d–f) Schematics of the magnetic-field directions corresponding to the measurements in (a–c).
(g) Device geometry used for magnetotransport measurements, indicating crystallographic axes and the orientation of current and voltage lines.
(h) Extracted amplitudes of the octupole-driven contribution $R^{\mathrm{Oct}}_{xy}$ (red) and the scalar-chirality contribution $R^{\mathrm{SSC}}_{xy}$ (black) as functions of in-plane angle. Both show an approximate 120° periodicity. Error bars represent the standard deviation of five repeated measurements.
(i) Symmetry-based theoretical calculation of scalar spin chirality, displaying the same 120° periodicity but with an opposite sign compared to the octupole contribution.
}

\end{document}